\documentclass[preprint]{aastex}
\usepackage{graphicx}
\newcommand{\eccactual}{$e = 0.07755^{+0.00022}_{-0.00027}$}
\newcommand{\wdotactual}{$\dot\omega = 0.07089^{+0.00021}_{-0.00013}$ deg~cycle$^{-1}$}

\newcommand{\eccvalue}{ $0.07755^{+0.00018}_{-0.00026}$} 
\newcommand{\wdotvalue}{$0.07089^{+0.00021}_{-0.00013}$}
\newcommand{\uvalue}{$33.48^{+0.10}_{-0.06}$}

\newcommand{\apsidalperiod}{$33.48^{+0.10}_{-0.06}$~yr}
\newcommand{\apsidalperiodheiser}{$30.4$~yr}
\newcommand{\methodoneapsidalperiod}{$33.06\pm0.58$~yr} 

\newcommand{\heiserwdot}{$\dot\omega = 0.0780$ deg~cycle$^{-1}$}
\newcommand{\methodonewdot}{$\dot\omega =0.0718\pm0.0012$ deg~cycle$^{-1}$} 

\newcommand{\Bone}{$104.00\pm0.26$} 
\newcommand{\Btwo}{$348.10\pm0.69$} 
\newcommand{\Bthree}{$140.8\pm1.6$} 

\newcommand{\persec}{\ensuremath{\mbox{s}^{-1}}}

\def\astrosun {\mbox{$\odot$}}
\newcommand{\Msol}{\ensuremath{\mbox{M}_{\astrosun}}}

\usepackage{natbib}
\usepackage{graphicx,epsfig}
\usepackage{bm}
\usepackage{color}
\usepackage{comment} 

\shorttitle{Apsidal Motion of V578 Mon} 
\shortauthors{Garcia et al.}

\begin{document}

\title{Apsidal Motion of the {\bf Massive}, Benchmark Eclipsing Binary V578 Mon}
 
\author{E.V.\ Garcia\altaffilmark{1,2}, Keivan G.\ Stassun\altaffilmark{3,1}, 
L.\ Hebb\altaffilmark{3}, Y.\ G\'omez Maqueo Chew\altaffilmark{3,4}, 
A.\ Heiser\altaffilmark{5}}

\altaffiltext{1}{Department of Physics, Fisk University,
1000 17th Ave.\ N., Nashville, TN, USA 37208; eugenio.v.garcia@gmail.com}
\altaffiltext{2}{Fisk-Vanderbilt Masters-to-PhD Bridge Program Graduate Fellow}
\altaffiltext{3}{Department of Physics \& Astronomy, 
Vanderbilt University, VU Station B 1807, Nashville, TN 37235, USA} 
\altaffiltext{4}{Department of Astronomy, Queens University Belfast, University Rd
Belfast, County Antrim BT7 1NN, United Kingdom}
\altaffiltext{5}{Vanderbilt Dyer Observatory, 1000 Oman Dr., Brentwood, 
TN 37027, USA} 

\begin{abstract} 
V578~Mon is a system of two early B-type stars in the Rosette Nebula 
star-forming region (NGC~2244), and is one of only nine eclipsing 
binaries with component masses greater than 10 \Msol\ whose physical 
parameters have been determined with an accuracy of better than 3\%. It is 
therefore a benchmark system for evolutionary and stellar structure models of newly formed massive 
stars. Combining our multi-band light curves spanning 40~yr with previous light
curve data from the literature, we fit a model light curve that for the first
time includes the effects of apsidal motion of the system. 
We measure an apsidal period of \apsidalperiod. 
As a consequence of incorporating the apsidal motion into the modeling of
the system's orbital parameters, we determine an
updated eccentricity of \eccactual, which differs significantly
from the value previously reported in the literature. 
Evidently, the inclusion of apsidal motion in the light curve modeling 
significantly affects the eccentricity determination.
Incorporating these key parameters into a comprehensive model of the system's
physical parameters---including internal structure constraints---will
bring V578~Mon to the next level of benchmark precision and utility.
\end{abstract}


\section{Introduction\label{sec:intro}} 

The analysis of apsidal motion in eccentric binary stars has been used
for years to test stellar structure models. 
The periastron
advance of an eccentric binary system is a direct consequence of the
finite size of the stellar components, and of tidal interactions between them. 
Consequently, apsidal motion serves as a measure of the internal structure of stars
\citep{1939MNRAS..99..451S}. 
Specifically, measurement of apsidal motion in eccentric
binary systems allow stringent tests of the internal structure constant
$k_{2}$ predicted by theory \citep[e.g.][]{2010A&A...519A..57C}.  Apsidal
motion also provides a test of General Relativity outside of our Solar
System \citep[e.g.][]{2010A&A...509A..18W}.

The photometric variability of the 2.408 day period, eccentric, massive eclipsing binary
(EB) V578~Mon (HDE~259135, BD$+4^\circ$1299), comprising a B1 type primary
star and a B2 type secondary
star, was first identified in the study by \citet{1977AJ.....82..973H} 
of NGC~2244 within the Rosette Nebula (NGC~2237, NGC~2246).  The absolute
dimensions of V578~Mon have been determined from three seasons
of Str\"{o}mgren $ubvy$ photometry and one season of radial-velocity
data by \citet[hereafter H2000]{2000A&A...358..553H}. An analysis of
the metallicity and evolutionary status of V578~Mon was undertaken by
 \citet{2005A&A...439..309P} and H2000.
The masses and radii of V578~Mon determined from these data are
$14.54\pm0.08$ \Msol\ and $10.29\pm0.06$ \Msol, and 
$5.23\pm0.06$ R$_{\odot}$ and
$4.32\pm0.07$ R$_{\odot}$, for the primary and secondary respectively (H2000).

These masses and radii are accurate to better than 3\%, making V578~Mon
one of only nine EBs with $M_1 \ge M_2 > 10$ \Msol\ and with 
sufficient accuracy to be included in the \citet{2010A&ARv..18...67T}
compilation of benchmark-grade EBs. However, of these, V578~Mon is the
only eccentric EB lacking an apsidal motion measurement.

The classical theory of tides and General Relativity both predict 
that a close, eccentric system such as V578~Mon will experience 
a certain amount of periastron advance. The angle of periastron, $\omega$,
is given by the equation $\omega(t) = \omega_{0} + \dot\omega t$, 
where $\omega_{0}$ is the angle of periastron at the reference epoch HJD$_{0}$, 
and $t$ is the time since HJD$_{0}$.
The apsidal period is given by $U = 360^\circ/\dot\omega \times P$,
where $P$ is the orbital period and $\dot\omega$ is the apsidal motion in 
deg~cycle$^{-1}$.  The AAVSO research note by \citet{2010JAVSO..38...93H} presented the 
first long term photometry of V578 Mon that was used to identify 
apsidal motion in the system and to update its ephemeris.  
The author estimated a value of $U \approx$~\apsidalperiodheiser~
from 14 independent primary eclipse minima measured over a 40 year period. 
Here we present an analysis of the apsidal motion of V578~Mon using the
state-of-the-art EB modeling software {\sc phoebe} \citep{2005ApJ...628..426P},
which updates and extends the venerable Wilson-Devinney code 
\citep{wilson1971,wilson1979}.
Traditionally, apsidal motion of EBs has been determined via eclipse timings 
\citep[e.g.][]{1992A&A...260..227G,1995Ap&SS.226...99G,2005A&A...437..545W}. 
However, the apsidal motion of an eccentric EB
causes not only the eclipse timings to vary; it also causes the
shapes and depths of the eclipses---as well as the morphology of the 
out-of-eclipse portions of the light curve---to vary over time. 
Therefore, in principle a full light curve model takes into account more
of the apsidal information encoded in the light curve data, and should yield
an extremely precise measure of the apsidal motion. 

Using our own light curve
data spanning 40~yr together with previous light curves from the literature,
we measure an apsidal period for V578~Mon of \apsidalperiod.
Furthermore, as a consequence of including the apsidal motion for the first
time into the analysis of the orbit of V578~Mon, we report an updated 
eccentricity of \eccactual, which differs from the previous
literature value of 0.0867$\pm$0.0006 (H2000).
These fundamental orbital parameters set the stage for follow-up analyses
to determine the internal structure of the stars in V578~Mon
for the first time, and to re-determine the stellar radii, 
for detailed tests of stellar evolution models with this benchmark system.

In \S\ref{sec:data}, we present the photometry used in this paper. 
In \S\ref{sec:analysis}, we perform the light curve analysis of all photometry. 
In \S\ref{sec:results}, we present the apsidal period and orbital 
eccentricity along with an error analysis. 
We conclude in \S\ref{sec:summary} with a discussion and summary.

\section{Data\label{sec:data}} 
The available time-series photometry of V578~Mon covers nearly 40~yr and 
more than one full apsidal motion period. 
A summary of the various light curve epochs, including filters and observing
facilities used, is presented in Table~\ref{Table:Phot}. 
Photometry from \citet{2010JAVSO..38...93H}
includes multiband light curves spanning 1967--2006 from the
16-in telescope at Kitt Peak National Observatory (KPNO) and from the
Tennessee State University(TSU) -Vanderbilt 16-in Automatic 
Photoelectric Telescope (APT) at Fairborn Observatory.  The KPNO
Johnson $UBV$ light curves comprise 725 data points spanning 1967--1984 
with average formal uncertainties per data point of 0.004 mag. 
The APT Johnson $BV$ light curves span 1994--2006 and consist
of 1783 data points with formal uncertainties per data point 
of 0.001 mag for B and 0.002 mag for V \citep{2010JAVSO..38...93H}. Light curves 
from H2000 span 1991--1994 from the 0.5-m Str\"{o}mgren Automatic Telescope (SAT) 
at La Silla, with 248 data points in each of the $ubvy$ filters and average formal 
uncertainty per data point of 0.003 mag (H2000). 
Table~\ref{Table:Phot} lists these formal uncertainties,
$\sigma_0$, as reported by the original authors. 
However, from our light curve fits (see below) we found that these 
formal errors were in most cases underestimated. Thus
we also report as $\sigma$ in Table~\ref{Table:Phot} the 
uncertainties that we ultimately adopted for each light curve
(see \S\ref{sec:globalfit} for details).

\section{Light Curve Analysis\label{sec:analysis}} 

Here we determine $\dot\omega$ with the light curve modeling program 
{\sc phoebe} using all of the available photometry (Table~\ref{Table:Phot}). 
We perform the light curve fitting in two separate approaches. 
In the first approach (\S\ref{sec:epochfit}) ,
we determined $\dot\omega$ by finding the linear change in $\omega(t)$ 
from fits to individual light curve epochs. 
In the second approach (\S\ref{sec:globalfit}), 
we determine $\dot\omega$ by finding a global light curve solution 
simultaneously to all of the light curve data in Table~\ref{Table:Phot}. 

All fixed parameters are listed in Table~\ref{Table:Params}. 
For fixed parameters in both approaches, we adopt the spectral type
of the primary to be B1V, which implies $T_1 = 30000$K (H2000,
and references therein).  We adopt gravity brightening ($g_1$, $g_2$)
and surface albedos ($A_1$, $A_2$) to be 1, as appropriate for 
stars with radiative envelopes. The rotational synchronicity parameters ($F_1$, $F_2$) are fixed at 
$1.13\pm0.03$ and $1.11\pm0.03$ for the primary and secondary, 
respectively, based on the $v\sin i$ and radii determined by H2000.
Our limb darkening coefficients follow the square-root law for fully 
radiative stars \citep{2000A&A...363.1081C}. In all cases we also adopt
the semi-major axis ($a$) and the mass ratio ($q\equiv M_2/M_1$) from H2000
because these parameters are determined principally from the radial-velocity
curves. We do not include star spots in any of the light curve modeling; 
as discussed below, the variations in the light curve data are fully
reproduced through the effects of apsidal motion.

\subsection{Fits to individual light curve epochs\label{sec:epochfit}} 
To obtain an initial, simple estimate of the apsidal period, we 
first determine $\dot\omega$ by finding the linear 
change in $w$ from the individual Johnson $BV$ light curve epochs 
1973--1976, 1994--1995, and 2005--2006 (Table~\ref{Table:Phot}). 
We chose these epochs because they span one full apsidal period, 
and because these light curves were obtained using the same instrument 
and filter set. Johnson $BV$ light curve epochs 1973--1976 are a portion 
of the KPNO 1967-1984 light curve listed in Table~\ref{Table:Phot}. 

The free parameters are the angle of periastron $w$,
the inclination of the orbit $i$, the secondary temperature $T_2$, and the
surface potentials $\Omega_1$ and $\Omega_2$.
The starting values for these parameters are from H2000 
(see Table~\ref{Table:Params}). 
Note that we fit for the surface potential of the stars (i.e., $\Omega \propto R^{-1}$)
as well as $T_2$, in order to give the light curve model full freedom to 
fit the eclipse widths and depths, however we regard the best-fit values
of these parameters as preliminary.
We fixed \eccactual, as determined in the global solution 
to all light curve data described in \S\ref{sec:globalfit}. 
We set the orbital period, $P$, to the value from \citet{2010JAVSO..38...93H},
and we used the primary minima eclipse
times from \citet{2010JAVSO..38...93H} as the HJD$_{0}$ values for each
light curve epoch. 

The resulting best-fit $\omega$ at specified HJD$_{0}$'s and formal 
errors from each light curve epoch are given in Table~\ref{Table:w}.
A least-squares line fit to these $\omega$ vs.\ orbital cycles initially gave 
a reduced chi square $\chi^2_{\rm red} =  \chi^2/N = 19.0$, 
where N is the number of data points. 
This indicates that the uncertainties on the individual
$\omega$ values are underestimated. This
is not surprising because the formal uncertainties on $\omega$ 
here do not take correlations with other parameters into account, as we do
in our global light curve solution below. Therefore we scaled the 
uncertainties in Table~\ref{Table:w} to achieve $\chi^2_{\rm red} = 1$.
The slope of the fitted line gives \methodonewdot\ and thus
an apsidal period $U =$~\methodoneapsidalperiod.

\subsection{Global fit to all light curve data\label{sec:globalfit}} 

Next we perform a global light curve solution fit simultaneously to all epochs 
and filters of available light curve data (see Table~\ref{Table:Phot}). 
The free parameters are $e$, $\omega_0$, $i$, $T_{2}$, $\Omega_{1}$, 
$\Omega_{2}$, and $\dot\omega$. The initial values for these parameters are 
from H2000 except for $\dot\omega$, for which we used
\heiserwdot\ calculated from the estimated apsidal period
from \citet{2010JAVSO..38...93H}. 
We again fixed $P$ and HJD$_0$ to the values from \citet{2010JAVSO..38...93H}. 
 
We perform two global light curve fitting iterations. For the first iteration 
we used the formal photometric uncertainties, $\sigma_0$ in 
Table~\ref{Table:Phot}, which yielded a total $\chi^{2}_{red}=3.25$,
indicating underestimated photometric errors.
Thus for the second iteration we scaled the photometric errors to make 
the $\chi^2$ of each light curve to equal the number of data points 
$N$ of the light curve.  The scaled photometric errors $\sigma$ listed in 
Table~\ref{Table:Phot} reflect more realistic, conservative 
values given the quality of our photometry.  
Effectively this causes $\chi^2_{red}$ of each light curve to be 
approximately unity.
This yielded a total $\chi^{2}_{red} = 0.97$
for the final fit. 

\begin{figure}[ht]
\epsscale{1.0}
\begin{center}
\includegraphics{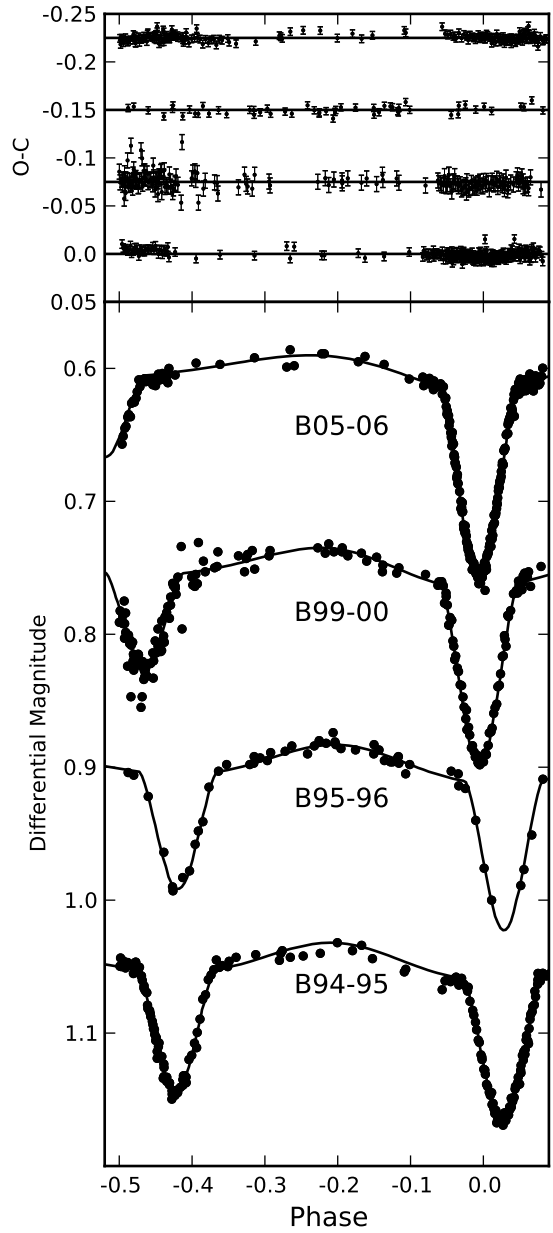}
\end{center}
\caption{\label{fig:B} Representative fits to light curves from 2005--2006, 
1999--2000, 1995--1996 and 1994--1995 in the Johnson $B$ passband from 
global fits to all light curve data, offset for clarity (see \S\ref{sec:globalfit}). 
The residuals to the fits $(O-C)$ are shown above. 
}
\end{figure}

\begin{figure}[ht]
\begin{center}
\includegraphics{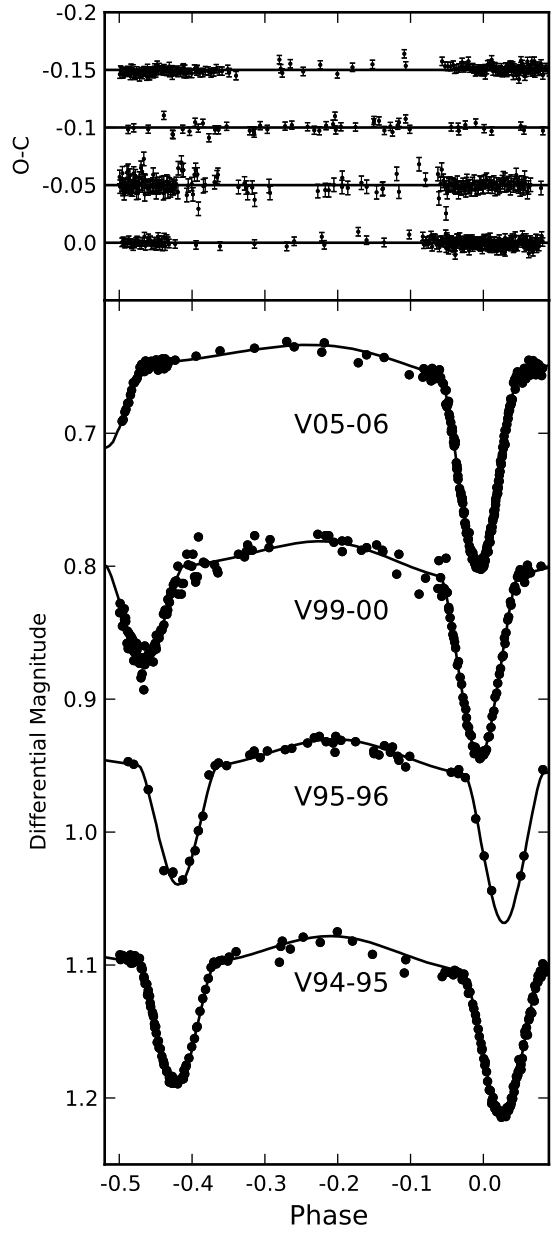}
\end{center}
\caption{\label{fig:V} 
Same as Fig.~\ref{fig:B}, but showing Johnson $V$ band light curves and fits.
}
\end{figure}

\begin{figure}[ht]
\begin{center}
\includegraphics{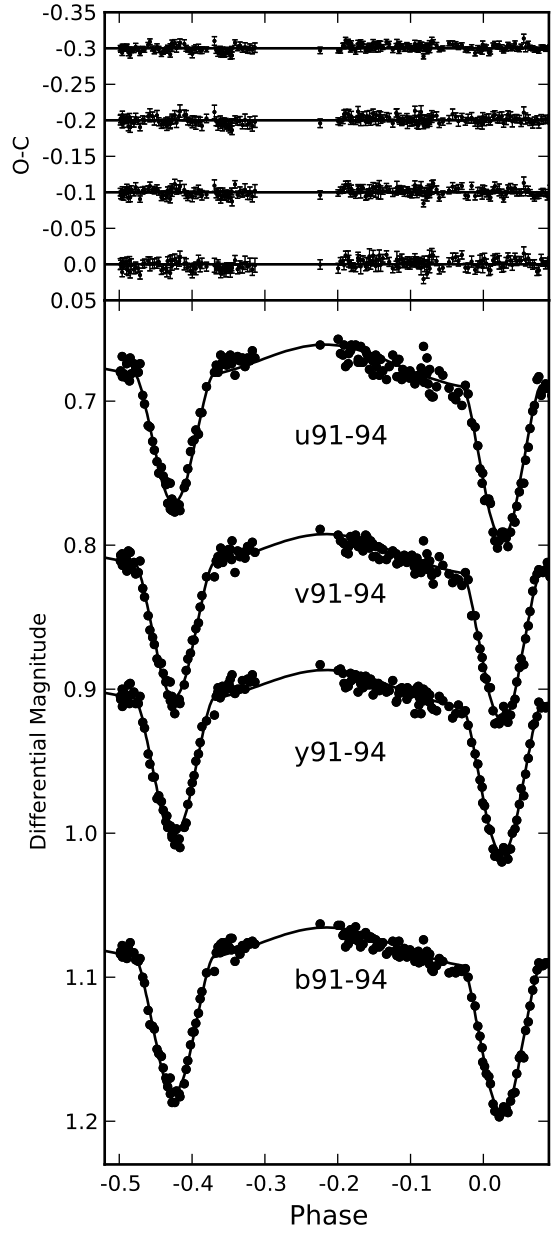}
\end{center}
\caption{\label{fig:SAT} 
Same as Fig.~\ref{fig:B}, but showing Str\"omgren $ubvy$ light curves and fits.
}
\end{figure}

\begin{figure}[ht]
\begin{center}
\includegraphics{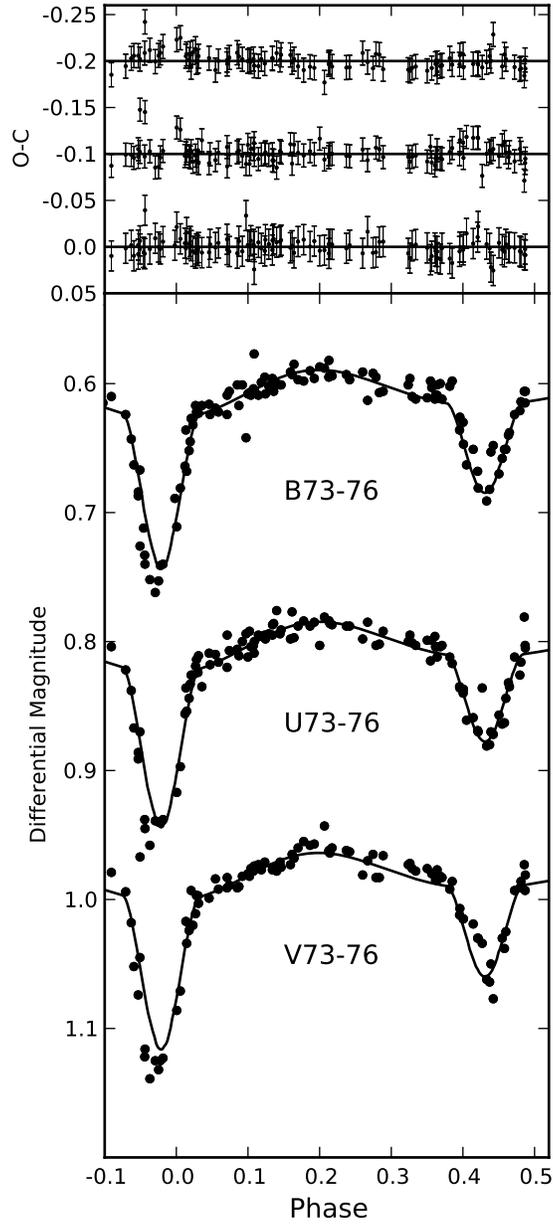}
\end{center}
\caption{\label{fig:KPNO} 
Same as Fig.~\ref{fig:B}, but showing 1973--1977 Johnson $UBV$ light 
curves and fits.
}
\end{figure}

The best fit global solution is plotted over multiple light curve epochs 
in Figures \ref{fig:B}, \ref{fig:V}, \ref{fig:SAT}, and \ref{fig:KPNO}. 
The resulting apsidal motion 
is \wdotactual, giving an apsidal period $U = $~\apsidalperiod.
The quoted uncertainties are from a 
detailed $\chi^2$ analysis (see \S\ref{sec:wdot_vs_e}). 
In Table~\ref{Table:Params}, we list the full set of system 
parameters resulting from the global light curve model fit.

\begin{figure}[ht]
\begin{center}
\includegraphics{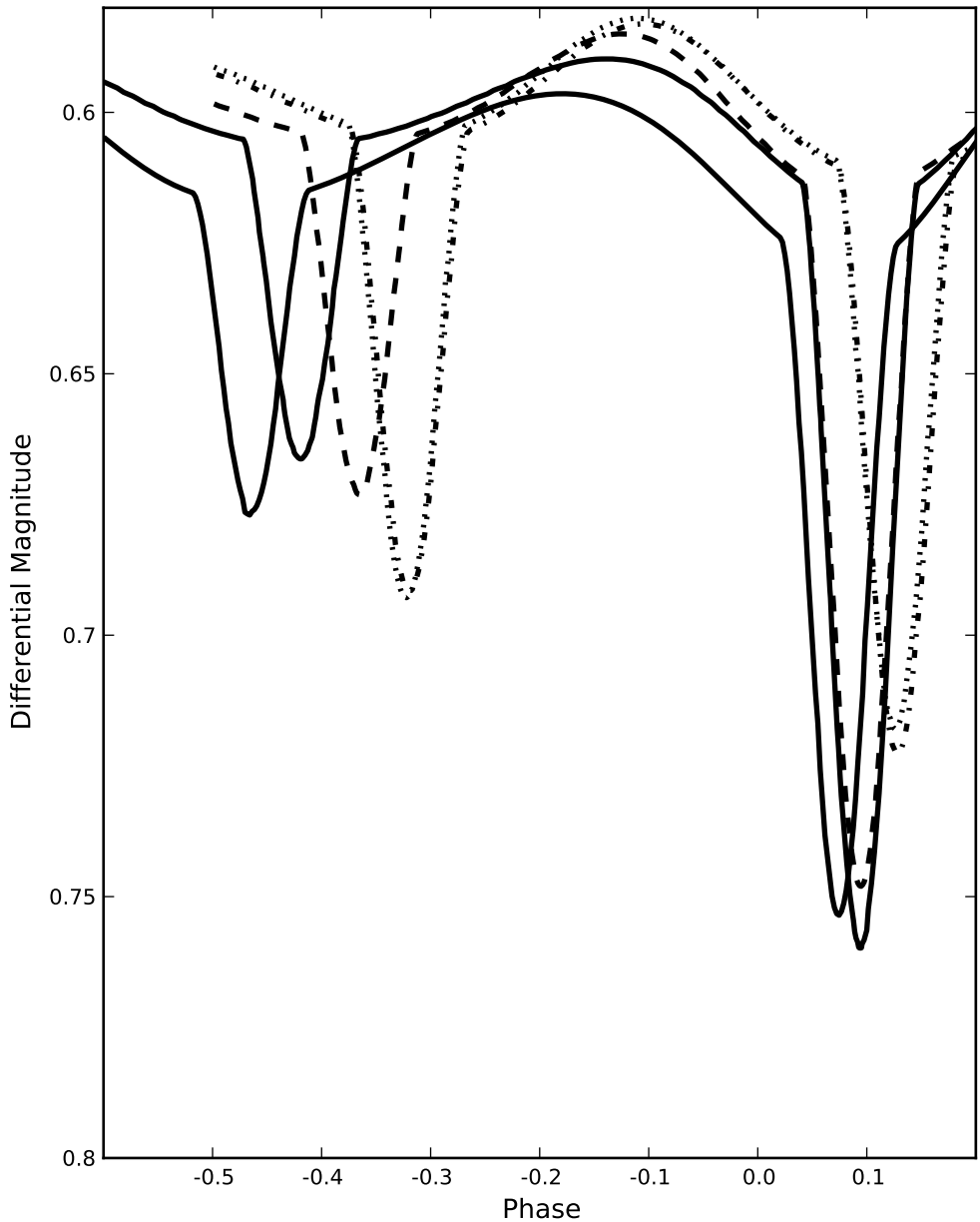}
\end{center}
\caption{\label{fig:demo} 
Representative light curve model fits
in the Johnson $B$ passband from our global light curve solution
(see \S\ref{sec:globalfit}). 
Note the variation of the shapes, depths, and
timing of the primary and secondary eclipses. The out-of-eclipse portions
of the light curves also vary, due to the apsidal motion, not star spots.
Solid lines represent fits to the 2005--2006 and 1973--1976 epochs.
The dashed, dash-dot, and dotted lines represent fits to
the 1999--2000, 1995--1996, and 1994--1995 epochs, respectively.
}
\end{figure}

Note that the changes in the shapes, depths, and timings
of the primary and secondary eclipses---as well as the changes in the
out-of-eclipse portions of the light curves---are due to apsidal 
motion effects and are very well reproduced by the model. These
variations over time due to apsidal motion are
clearly demonstrated in Figure~\ref{fig:demo}, which displays the global light
curve solution for the Johnson $B$ light curve epochs 
2005--2006, 1999--2000, 1995--1996, 1994--1995, and 1973--1976. 

\section{Results\label{sec:results}} 

\subsection{Apsidal period of V578~Mon}

We have determined the apsidal motion, $\dot\omega$, of V578~Mon via model
fits to the available light curve data. 
Calculating the linear change in $\omega(t)$ by fitting individual
light curve epochs spaced over one apsidal period results
in \methodonewdot\ and $U = $~\methodoneapsidalperiod.
Calculating $\dot\omega$ via a global light
curve solution results in \wdotactual\ and $U = $~\apsidalperiod.
The two values are consistent within the uncertainties.
We adopt the latter value as it incorporates the full dataset, is more 
precise, and furthermore our error analysis accounts for correlations in 
the fitted parameters (see below).

We prefer the approach of fitting the full light curves
because it incorporates all available light curve 
data in Table~\ref{Table:Phot}, not just the eclipse timings.  As shown
across multiple observing seasons in 
Figures \ref{fig:B}, \ref{fig:V}, and \ref{fig:KPNO}, the shapes,
depths, and timings of the primary and secondary eclipses change with time
due to $\dot\omega$. Furthermore, the light curves show variation in
the out-of-eclipse data over time which are very well reproduced by the 
light curve model. The model light curves do not include star spots, 
showing that the out-of-eclipse variations in the light curve of V578~Mon 
are indeed a manifestation of apsidal motion. 

\subsection{Orbital eccentricity of V578~Mon} 

The eccentricity from our light curve model fit incorporating 
apsidal motion is \eccactual\ (Table~\ref{Table:Params}),
which differs significantly
from the previously reported value of $0.0867 \pm 0.0006$ (H2000). 

To further investigate the H2000 eccentricity, we perform a light curve fit within 
{\sc phoebe} using only the SAT photometry (the same data used in H2000). 
We set $\dot\omega$ to zero and $a$, $q$, $w_0$, $i$, $T_{2}$, $\Omega_{1}$ 
and $\Omega_{2}$ to values from H2000. The only free parameter is $e$. 
This {\sc phoebe} fit converges to a light curve solution with
$e = 0.0867$, reproducing the $e$ found by H2000. 
Evidently, accounting for the effects of apsidal motion yields a different $e$.
The $e$, $\Omega_{1}$ and $\Omega_{2}$ are correlated parameters in 
light curve analysis, meaning that our significantly different $e$ 
could yield different radii for the stars.  The new, tentative radii we compute 
are approximately $5.14$ R$_{\odot}$ and $4.70$ R$_{\odot}$ for the primary 
and secondary as compared to the literature values of $5.23\pm0.06$ R$_{\odot}$ and
$4.32\pm0.07$ R$_{\odot}$ (H2000). Thus there is an indication that the updated $e$ together with 
the $\dot\omega$ newly reported here may result in a significantly different R$_{2}$.
This will be the subject of an in-depth analysis in a forthcoming paper.

\subsection{Uncertainties on $\dot\omega$ and $e$ \label{sec:wdot_vs_e}}
In order to determine realistic uncertainties on $\dot\omega$ and $e$, 
we performed a detailed analysis of the $\chi^2$ space around the best
fit values. 
We varied $\dot\omega$ over 0.0702--0.0716 deg~cycle$^{-1}$ 
with step length 0.00006 deg~cycle$^{-1}$, and we varied $e$ over
0.0765--0.0787 with step length 0.000084.
For each of the 625 combinations of $\dot\omega$ and $e$, 
we recomputed the global light curve fit as before.

\begin{figure}[ht]
\begin{center}
\includegraphics[width=1.1\textwidth,height=5in]{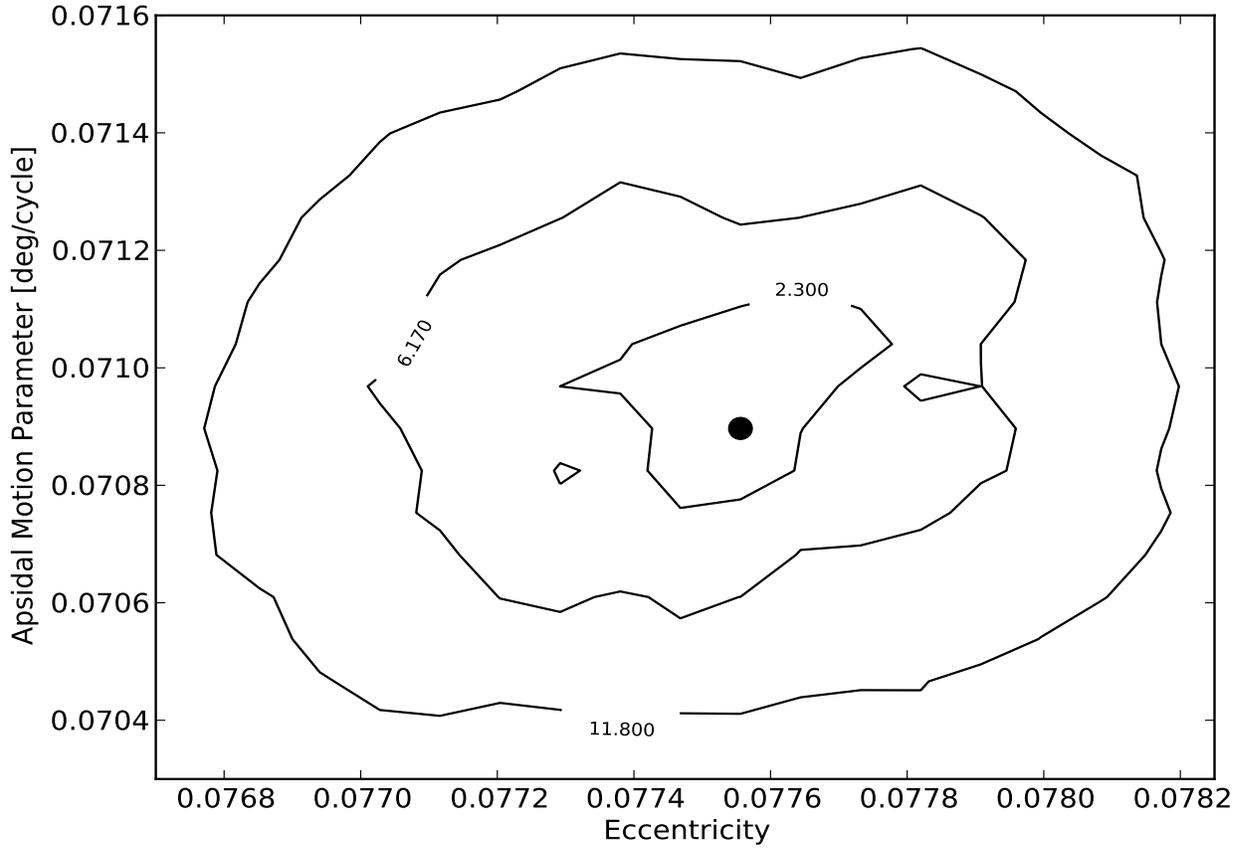}
\end{center}
\caption{\label{fig:wdot_vs_e} 
Constraints on the apsidal motion $\dot\omega$ and eccentricity $e$ described
in \S\ref{sec:wdot_vs_e}. The contours represent 
$\Delta\chi^2 = \chi^2 - \chi^2_{min}$ 
values of 2.30, 6.170, and 11.80 (corresponding to 
1, 2, and 3 $\sigma$ confidence intervals).
The black dot represents our best fit solution.
The resulting best fit and uncertainties are \wdotactual\ and \eccactual.
}
\end{figure}

Figure~\ref{fig:wdot_vs_e} shows the resulting contour plot of $\chi^2$ 
for $e$ vs.\ $\dot\omega$. Contours are drawn at $\Delta\chi^2$ values
corresponding to 1$\sigma$, 2$\sigma$, and 3$\sigma$ uncertainty
for a $\Delta\chi^{2}$ distribution of two parameters of interest \citep{NR}. 
The contour shapes indicate that $\dot\omega$ and $e$ are not strongly
correlated given our analysis. Moreover, Figure~\ref{fig:wdot_vs_e} clearly demonstrates that
the previously reported value of $e = 0.0867 \pm 0.0006$ (H2000)
lies well beyond the 3$\sigma$ contour and can therefore be ruled out
with very high statistical significance.

\section{Discussion and Summary\label{sec:summary}} 

The accurate apsidal period of \apsidalperiod\ and updated orbital
eccentricity of \eccactual\ of V578~Mon underscores the value of a
long time baseline of photometric observations for eccentric eclipsing 
binary stars. The traditional eclipse timing method \citep[e.g.][]{1995Ap&SS.226...99G}
uses the timing of the primary
and secondary eclipse of an eccentric binary star system to calculate $e$
and $\dot\omega$ among other parameters. In this paper, we use the eclipse
timings but also the changing shapes, widths, and depths of the primary and
secondary eclipse due to $\dot\omega$, as well as the light curve variations
in the out-of-eclipse phases to determine realistic constraints on $\dot\omega$ and $e$. 
Furthermore, we demonstrate that including the apsidal motion parameter in light curve fitting 
can affect the eccentricity measurement of eclipsing binary systems.

The apsidal motion of V578~Mon can provide an accurate test of theoretical
calculations of the internal structure constant ($k_{2}$). For example,
\citet{2010A&A...519A..57C} find generally good agreement between the 
theoretically predicted and measured $k_2$ when they consider EBs
with radii to $\pm$2\% accuracy. Comparing against EBs with very accurately
measured radii is critical, because the theoretically predicted $k_2$ is 
highly dependent upon the stellar radii ($k_2 \propto R^5$). 
Currently, only 18 eccentric EBs have stellar radii and $\dot\omega$ 
measured with sufficient accuracy to stringently test theoretical
internal structure models. Of these, only EM~Car, V478~Cyg, 
V453~Cyg, and CW~Cep have masses greater than 10~\Msol, 
well constrained $\dot\omega$, and radii accurate to $\pm$2\%
\citep[][and references therein]{2010A&A...519A..57C}.
A detailed reanalysis of V578~Mon will yield accurate radii and,
combined with our \wdotactual, will yield an accurate $k_2$
with which to test theoretical internal structure models of high mass stars.

There is still much to know about high mass stellar evolution, especially at young ages, 
making V578~Mon an important testbed for stellar evolution models given the large amount of
precise photometry and radial velocity data on the system. To date, there
are only nine EBs with component masses greater than 
10 \Msol\ whose masses and radii are accurate to $\pm$3\%
\citep{2010A&ARv..18...67T}.  Stellar evolution models for stars with masses
greater than 10 \Msol\ thus remain poorly constrained by EBs.
Future reanalysis of V578~Mon will include
a precise calibration of high-mass evolution models similar to the work
on V453~Cyg by \citet{2004MNRAS.351.1277S}. Finally, V578~Mon's location
allows for precise age and distance constraints of the Rosette Nebula,
similar to previous work on V578~Mon by H2000.
A follow-up paper will incorporate the accurate orbital parameters newly 
determined here in order to re-determine all of the physical 
properties---including internal structure parameters---of this important, 
benchmark EB.

\acknowledgments
We acknowledge the support of the NSF REU program in Physics \& Astronomy
at Vanderbilt University, and NSF grants AST-0849736 and AST-1009810.

\begin{deluxetable}{lllllllll}
\tablecolumns{7}
\tablewidth{0pt} 
\tablecaption{\label{Table:Phot} V578 Mon Light Curves}
\tablehead{ 
\colhead{Observatory} & \colhead{Year} & \colhead{Filter}& \colhead{$\sigma_{0}$} & \colhead{$\sigma$} & \colhead{N}  \\
\colhead{} & \colhead{} & \colhead{} & \colhead{(mag)} &\colhead{(mag)} & \colhead{} 
}
\startdata 
$^{1}$ KPNO & 1967-84 & Johnson $U$ &  0.004 & $0.016 $& 251\\ \hline 
& & Johnson $B$ & 0.004 & $0.012 $ & 256 \\ \hline
& & Johnson $V$ &  0.004 & $0.013 $ & 217 \\ \hline
$^{2}$SAT & 1991--94& Str\"{o}mgren $u$ & 0.0029 & 0.0067 & 248 \\ \hline
& & Str\"{o}mgren $b$ &  0.0023 &  0.0046& 248 \\ \hline
& & Str\"{o}mgren $v$ &  0.0023 &  0.0054 & 248 \\ \hline
& & Str\"{o}mgren $y$ &  0.0030 &  0.0053 & 248  \\ \hline
$^{3}$APT & 1994--95 & Johnson $V$ & 0.0037 & $0.0022$ &  260 \\ \hline
&  & Johnson $B$ &  0.001 &  $0.0040$ &  254 \\ \hline
 APT & 1995--96 & Johnson $V$ & 0.002 &  $0.0035 $ & 95  \\ \hline
&  & Johnson $B$ & 0.001 & $0.0037 $ & 96  \\ \hline
APT &  1999--2000& Johnson $V$ & 0.002 & $0.0058 $ & 259 \\ \hline
& & Johnson $B$ & 0.001 & $0.0078 $ & 246  \\ \hline
APT & 2005--06 & Johnson $V$ & 0.002 & $0.0036$ & 284  \\ \hline
& & Johnson $B$ & 0.001 & $0.0044$  & 283 \\ \hline
\enddata
\tablecomments{ $^{1}$16-inch telescope at Kitt Peak (KPNO)\\
$^{2}$0.5 m telescope at La Silla (SAT)\\
$^{3}$TSU-Vanderbilt 16-inch telescope at Fairborn University (APT)\\}
\end{deluxetable}

\begin{deluxetable}{lrrrc}
\tablecolumns{3}
\tablewidth{0pc} 
\tablecaption{\label{Table:Params} 
V578 Mon Parameters
}
\tablehead{} 
\startdata
\colhead{Free Parameters} & \colhead{H2000 Value} & \colhead{This study} \\ \hline
Apsidal Motion, $\dot\omega$ [deg~cycle$^{-1}$] & ... & \wdotvalue \\ \hline 
Apsidal Period, $U$ [yr] & ... & \uvalue \\ \hline
Angle of Periastron, $w_{0}$ [deg] & $153.3\pm0.3$ & $159.8\pm0.33$ \\ \hline 
Eccentricity, $e$ & $0.0867\pm 0.0006$ & \eccvalue \\ \hline
Inclination, $i$ [deg]  & $72.58 \pm 0.30$ & $71.67$ * \\ \hline 
Primary Surface Potential, $\Omega_{1}$ & $5.02\pm .05$ & $5.15$ * \\ \hline 
Secondary Surface Potential, $\Omega_{2}$ & $4.87\pm .06$ & $4.56$ *\\ \hline  
Secondary Temperature, $T_{2}$ [K] & $26400\pm400$ & 26100 * & \\ \hline 
\\ \hline
\colhead{Fixed Parameters} \\ \hline
Ephemeris, HJD$_{0}$ [d] & &$ 2449360.6250$ & \\ \hline
Orbital Period, P [d] & &$ 2.4084822$& \\ \hline 
Primary Surface Temperature, $T_{1}$ [K] & & $30000\pm 500$ &\\ \hline 
Systemic Velocity, $\gamma$ [km\persec] & & $34.9\pm0.1$ & \\ \hline
Semi-major Axis, $a$ [$R_{\odot}$] &  & $22.03\pm0.04$ &\\ \hline 
Mass Ratio, $q = M_{2}/M_{1}$ & & $0.7078 \pm .0002$ &\\ \hline 
Primary Synchronicity Parameter, $F_1$ & & $1.13\pm0.03$ & \\ \hline 
Secondary Synchronicity Parameter, $F_2$ & & $1.11\pm0.03$ & \\ \hline
\\ \hline
\enddata
\tablecomments{* These parameters are preliminary. }
\end{deluxetable}

\begin{deluxetable}{lrrr}
\tablecolumns{4}
\tablewidth{0pt} 
\tablecaption{\label{Table:w} $\omega(t)$ values from fitting individual light curve epochs}
\tablehead{ 
\colhead{Light Curve} & \colhead{$HJD_{0}$} & \colhead{$\omega$} & \\
\colhead{ } &  \colhead{[days]} & \colhead{[deg]} 
}
\startdata 
Johnson $BV$ 2005--06 & 2453741.73648 & \Bone \\ \hline 
Johnson $BV$ 1994--95 & 2449738.75624 &\Btwo \\ \hline
Johnson $BV$ 1973--77 & 2443105.90927 & \Bthree \\ \hline
\enddata
\end{deluxetable}

\begin{figure}[ht]
\epsscale{1.0}
\begin{center}
\includegraphics[width=\textwidth,height=5in]{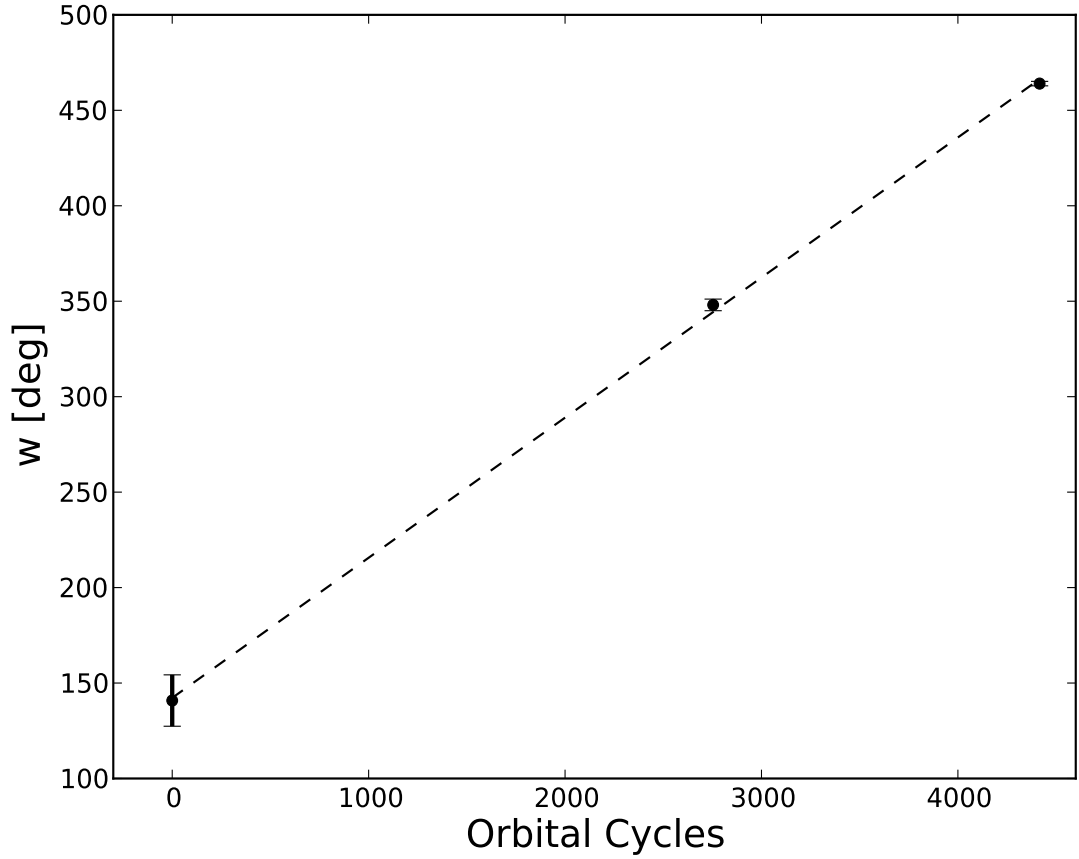}
\end{center}
\caption{\label{fig:ap} Variation of angle of periastron $\omega$ determined
by model fits to individual light curve epochs.
The circles represent $\omega$ values as determined from light
curves at the epochs 1974--1977, 1994--1995, and 2005--2006 (Table
\ref{Table:w}). Uncertainties on the individual $\omega$ are smaller than the
plotting symbols.
The dashed line represents a linear fit using the equation
$\omega(t) = \omega_{0} + \dot\omega t$, 
where $t$ is time since the reference epoch, HJD$_0$.  
This results in \methodonewdot, and $U = $\methodoneapsidalperiod.}
\end{figure}

\begin{thebibliography}{}
\bibitem[Claret \& Gim{\'e}nez(2010)]{2010A&A...519A..57C} Claret, A., \& Gim{\'e}nez, A.\ 2010, \aap, 519, A57 
\bibitem[Claret(2000)]{2000A&A...363.1081C} Claret, A.\ 2000, \aap, 363, 1081 
\bibitem[Gim{\'e}nez \& Bastero(1995)]{1995Ap&SS.226...99G} Gim{\'e}nez, A., \& Bastero, M.\ 1995, \apss, 226, 99 
\bibitem[Gimenez \& Quintana(1992)]{1992A&A...260..227G} Gimenez, A., \& Quintana, J.~M.\ 1992, \aap, 260, 227 
\bibitem[Heiser(2010)]{2010JAVSO..38...93H} Heiser, A.~M.\ 2010, Journal of the American Association of Variable Star Observers, 38, 93 
\bibitem[Heiser(1977)]{1977AJ.....82..973H} Heiser, A.~M.\ 1977, \aj, 82, 973 
\bibitem[Hensberge et al.(2000)]{2000A&A...358..553H} Hensberge, H., Pavlovski, K., \& Verschueren, W.\ 2000, \aap, 358, 553 
\bibitem[Pavlovski \& Hensberge(2005)]{2005A&A...439..309P} Pavlovski, K., \& Hensberge, H.\ 2005, \aap, 439, 309 
\bibitem[Pr{\v s}a \& Zwitter(2005)]{2005ApJ...628..426P} Pr{\v s}a, A., \& Zwitter, T.\ 2005, \apj, 628, 426 
\bibitem[Press(1988)]{NR} Press, W. H., Teukolsky, S. A., Vetterling, W. T. and Flannery, B. P., 1988, Numerical Recipes in C, Second Edition, Cambridge University Press, NY, pg 697. 
\bibitem[Southworth et al.(2004)]{2004MNRAS.351.1277S} Southworth, J., Maxted, P.~F.~L., \& Smalley, B.\ 2004, \mnras, 351, 1277
\bibitem[Sterne(1939)]{1939MNRAS..99..451S} Sterne, T.~E.\ 1939, \mnras, 99, 451 
\bibitem[Torres et al.(2010)]{2010A&ARv..18...67T} Torres, G., Andersen, J., \& Gim{\'e}nez, A.\ 2010, \aapr, 18, 67 
\bibitem[Wilson \& Devinney(1971)]{wilson1971} Wilson, R.~E.~\& Devinney, E.~J.~1971, \apj, 166, 605
\bibitem[Wilson(1979)]{wilson1979} Wilson, R.~E.\ 1979, \apj, 234, 1054
 

\bibitem[Wolf et al.(2010)]{2010A&A...509A..18W} Wolf, M., Claret, A., Kotkov{\'a}, L., Ku{\v c}{\'a}kov{\'a}, H., Koci{\'a}n, R., Br{\'a}t, L., Svoboda, P., \& {\v S}melcer, L.\ 2010, \aap, 509, A18 
\bibitem[Wolf \& Zejda(2005)]{2005A&A...437..545W} Wolf, M., \& Zejda, M.\ 2005, \aap, 437, 545 

\end{thebibliography}
\end{document}